\documentclass[11pt]{article}
\usepackage{SOTA_style}

\usepackage{lipsum}  

\usepackage{multirow}
\usepackage{diagbox}
\usepackage{xfrac}
\usepackage{tikz}
\usepackage{caption, subcaption}
\usepackage{pgfplots}
\usepackage{lscape}
\usepackage{pgfplots}
\usepackage[export]{adjustbox}
\usepackage{xcolor}

\title{State of the Art: Content-based and Hybrid Phishing Detection
}

\author{Luis Felipe Castaño\\
    \href{mailto:felipe.castano@unileon.es}{\texttt{felipe.castano@unileon.es}} 
\and Eduardo Fidalgo\\
    \href{mailto:efidf@unileon.es}{\texttt{eduardo.fidalgo@unileon.es}} 
\and Enrique Alegre\\
    \href{mailto:ealeg@unileon.es}{\texttt{enrique.alegre@unileon.es}} 
\and Deisy Chaves\\
    \href{mailto:dchas@unileon.es}{\texttt{deisy.chaves@unileon.es}} 
\and Manuel Sánchez-Paniagua\\
    \href{mailto:msancp@unileon.es}{\texttt{manuel.sanchez@unileon.es}}
    }
    
\begin{document}
{\setstretch{.8}
\maketitle
\begin{abstract}

    Phishing attacks have evolved and increased over time and, for this reason, the task of distinguishing between a legitimate site and a phishing site is more and more difficult, fooling even the most expert users. The main proposals focused on addressing this problem can be divided into four approaches: List-based, URL based,  content-based, and hybrid. In this state of the art, the most recent techniques using web content-based and hybrid approaches for Phishing Detection are reviewed and compared.

\noindent

\end{abstract}
}


\section{Introduction}

Phishing is a crime that uses social engineering to steal consumers’ personal identity data and financial account credentials, taking advantage of unsuspecting victims by making them believe that they are dealing with a legitimate and trusted party, through the use of deceptive email addresses, and email messages. These are designed to lead consumers to Websites that trick recipients into divulging financial data such as usernames and passwords. Phishing attacks have evolved and increased over time, Anti-Phishing Working Group (APWG) report phishing attacks rise in the third quarter of 2020, \textbf{detecting around 200,000 phishing sites in September}, APWG also informed that from 2016 until the third quarter of 2020 \textbf{phishing attacks hosted on HTTPS have increased from 10\% to 80\%}  \cite{Anti-PhishingWorkingGroup2020}, fooling even the most expert users.

The first approach to detect phishing websites was using the URL,  being able to classify websites even if they were not reachable, but in the same way as phishing techniques evolved, the techniques used to detect them did too. Currently, the approaches for detecting phishing webpages can be classified as follows: List-based, URL based, content-based, hybrid, and images-based.

\textbf{List-based approaches} can be applied in two ways: the first one is allowing only sites registered in a white-list as proposed Jain et al. \cite{Jain2016} in 2016; the second is restricting the sites that appear on a black-list \cite{Prakash2010}. However, these solutions need to be constantly updated and the attacker can bypass them just by making small changes to the URL.

\textbf{URL-based approaches} analyze exhaustively the URL components (protocol, domain, TLD, path) looking for variations in order to distinguish between a legitimate website and a phishing website. In these approaches, there are proposals such as \cite{Sahingoz2019, Moghimi2016, Somesha2020, Zhu2019}. These methods currently compare legitimate landing webpage with reported phishing webpage for training and testing their results. However, recently Sanchez-Paniagua et al. \cite{Paniagua19} proposed to compare legitimate login webpages with reported phishing, due this corresponds with the real problem:  determine if a login form of a website is legitimate or phishing. 

\textbf{Content-based approaches} use website content, such as HTML, CSS, and Javascript as input for algorithms, either raw data or as a vector of features extracted from a NLP technique to identify patterns between phishing websites. On the other hand, \textbf{hybrid approaches} use both  URL and content-based data for the same purpose. These two approaches will be deepened in the next section.

\textbf{Images-based approaches} compare screenshots of both legitimate and phishing webpage to compute the visual similarity. Among the techniques in this type of approach, it can be found the method proposed by Gangwar et al. \cite{Gangwar18} which extracts a fingerprint from the website that can be compared with a previously collected phishing fingerprint  dataset.

This document is organized as follows. We present a revision about the state of the art Content-based and Hybrid Phishing Detection, in Section \ref{sec:ContentbasedandHybrid}. Finally, we draw conclusions in Section \ref{sec:Conclusions}.

\section{Literature Review}
\label{sec:ContentbasedandHybrid}
Content-based or hybrid methods can be divided into two groups. The first one corresponds with \textbf{Automatic Features} where the raw data is used directly into an artificial intelligence algorithm, leaving the job of determining the important features on the algorithm. The second group are methods that use \textbf{Handcrafted Features}. They obtain a feature vector by using an NLP technique and later on this vector is the input for a classification algorithm.

\subsection{Automatic Features}
\label{subsec:AutomaticFeatures}
Opara et al. \cite{Opara2019} proposed the use of characters embedding and string embedding techniques to represent features of each HTML, then this representation is used as input to a Convolutional Neural Network (CNN) in order to model semantic dependencies. They collect their own data from Alexa and Phishtank, reporting two sets of data, the first one with 23000 legitimate websites and 2300 phishing websites used for training, and the second one with 24000 legitimate websites and 2400 phishing websites used for testing, these datasets are not available. Then, they obtained an accuracy of 98.00\% and an F1 score of 97.00\%. Finally, they affirm that automatic feature selection is the solution to the current approaches problems because the accuracy of existing models depends on how comprehensive the feature set is and how robust the feature set remains to future attacks, and also because the handcrafted approach requires substantial feature engineering.

Recently, Opara et al. \cite{Opara2020} proposed a technique that also uses automatic features, this time using the URL and HTML embeddings as input to a deep learning algorithm. The URL and HTML strings are tokenized using a character corpus that includes punctuation marks, then, this tokenized data is processed into a character embedding matrix. They use the datasets presented in their previous work, \cite{Opara2019}, reporting an accuracy of 98.00\% and an F1 score of 98.00\%.

\subsection{Handcrafted Features}
Adebowale et al. \cite{Adebowale2019} proposed an Adaptive Neuro-Fuzzy Inference System trained from images, frames, and text of the webpage, extracting 35 features. They used The University of Huddersfield  \cite{Mohammad2015} web phishing public dataset for training and testing, achieving 98.30\% of accuracy and 98.28\% of F1 score.

Yang et al. \cite{Yang2019}, proposed a stack of algorithms for the detection of phishing. First, a CNN-LSTM algorithm for detecting phishing using URL features. Then, the result of the CNN-LSTM algorithm was joined to webpage code features and webpage text features. And finally,  all these features are used by XGBoost algorithm for the final classification, achieving an accuracy of 99.41\% and F1 score of 99.00\%. They Collect their own data from PhishTank and dmoztools.net, using 1021758 phishing webpages and 989021 legitimate webpages, this dataset was not released.

Rao et al. \cite{Rao2019} implemented a system using three types of features: URL Obfuscation features, Third-Party-based features and Hyperlink-based features. These features are the input for a Random Forest algorithm that achieves 99.55\% accuracy using data collected from Alexa and Phishtank with 1407 legitimate webpages and 2119 phishing webpages.

Ozker et al. \cite{Ozker2020} studied the use of multiple machine learning algorithms. They identified 58 different features in the HTML content and then applied several machine learning methods like Naive Bayes, Random Forest, Support Vector Machine, Logistic Regression, K-Nearest Neighbors, Decision Tree, Multilayer Perceptron, and XGBoost. Random Forest showed better results with an accuracy of 97.91\% and F1 score of 98.00\%. They Collect their own data from PhishTank, using 8353 phishing webpages and 5438 legitimate webpages, this dataset was not released.

Li et al. \cite{Li2019} proposed a two-layer stacking algorithm where the first layer consists of three basic models: Gradient Boosting Decision Tree (GBDT), XGBoost, and LightGBM, using a strategy similar to K-fold cross-validation to train the basic models. Then, they combine the first input features and the results of the first layer of the stacking models as the final features. This final features are used to train a GBDT model to make predictions on the phishing webpages. For the tests, they collect and release a dataset called 50K Image Phishing Detection Dataset, reporting accuracy of 97.30\%. It was not possible to find the dataset at the moment of this review although according to the authors it was released.

Alotaibi et al. \cite{Alotaibi2020} proposed the use of a voting algorithm for feature selection. Then they used AdaBoost and LightGBM ensemble methods to detect phishing websites. They used two public datasets for testing, "Phishing Dataset for Machine Learning: Feature Evaluation"  \cite{Tan2018} and "Phishing Websites Data Set" \cite{Mohammad2015}, reporting an accuracy of 97.05\% and an F1 score of 97.35\%.

\begin{table}[phtb!]
\begin{center}
\resizebox{\linewidth}{!} {
\begin{tabular}{ l c c c c c c c c}
\hline
\multicolumn{9}{c}{\textbf{Automatic Features}}\\
\hline
\multirow{2}{*}{Method} & \multirow{2}{*}{Type} &\multirow{2}{*}{Technique} &\multirow{2}{*}{Year}
&\multicolumn{3}{c}{Dataset}
&\multicolumn{2}{c}{Results (\%)}\\
\cline{5-9}
 & & & & Name & N Legitimate & N Phishing & Accuracy & F1 Score\\
\hline
Opara et al. \cite{Opara2019}& Content & CNN & 2019 & Collected data & 24,000 & 2,400 & 98.00 & 97.00\\
Opara et al. \cite{Opara2020} & Hybrid & Deep learning & 2020 & Collected data & 47,000 & 4,700 & \textbf{98.00} & \textbf{98.00}\\
\hline
\multicolumn{8}{c}{\textbf{Handcrafted Features}}\\
\hline
\multirow{2}{*}{Method} & \multirow{2}{*}{Type} &\multirow{2}{*}{Technique} &\multirow{2}{*}{Year}
&\multicolumn{3}{c}{Dataset}
&\multicolumn{2}{c}{Results (\%)}\\
\cline{5-9}
 & & & & Name & N Legitimate & N Phishing & Accuracy & F1 Score\\
\hline
Adebowale et al. \cite{Adebowale2019} & Hybrid & Neuro-Fuzzy Inference & 2019 &  \cite{Mohammad2015} & 7,262 & 3,793 & 98.30 & 98.28\\

Yang et al. \cite{Yang2019} & Hybrid & CNN - LSTM - XGBoost & 2019 &  Collected data & 22,390 & 22,445 & \textbf{99.41} & \textbf{99.00}\\

Rao et al. \cite{Rao2019} & Hybrid & Random Forest & 2019 &  Collected data & 1,407	 & 2,119 & \textbf{99.55} & -\\

Ozker et al. \cite{Ozker2020} & Content & Random Forest & 2020 &  Collected data & 5,438 & 8,353 & 97.91 & 98.00\\

Li et al. \cite{Li2019} & Hybrid & Stacking model & 2019 &  50K PD\cite{Li2019}   & 28,320 & 24,789 & 97.30 & -\\

Alotaibi et al. \cite{Alotaibi2020} & Hybrid & AdaBoost and LightGBM & 2020 &  \cite{Tan2018} - \cite{Mohammad2015}  & 5,000/7,262 & 5,000/3,793 & 97.05  & 97.35\\

\hline

\end{tabular}
}
\caption{\label{tab:literatureReview} 
Comparative table of the methods presented in section \ref{sec:ContentbasedandHybrid} - Content-based and Hybrid Approaches}
\end{center}
\end{table} 

\section{Discussion and Conclusions}
 \label{sec:Conclusions}
Although the best reported result is presented in the method proposed by Rao et al. \cite{Rao2019}, in the table summary \ref{tab:literatureReview} it can be seen that the dataset used for testing is small compared to other methods and also they use third-party features, making it dependent on external services. The method proposed by Yang et al. \cite{Yang2019} is the one that reports the second-highest accuracy and the best F1 score value. It was trained with the largest and most balanced dataset reported in the reviewed methods. However, this method uses third-party features and the complete list of selected features is not clear enough. On other hand, the technique proposed by Opara et al. \cite{Opara2020} in the section of automatic features \ref{subsec:AutomaticFeatures}  shows good results, their method was trained with an unbalanced dataset as can be seen in table summary \ref{tab:literatureReview}, which can easily be improved.

Most of the works reviewed use a hybrid approach, and some of them compare the results obtained using only the content against the achieved using both, reporting an increase in the performance of the algorithm when using both, HTML and URL content. This is why we think using a hybrid method is the better option. Due to the above and the changing nature of phishing, we have decided to explore hybrid techniques with Automatic Features as shown in section \ref{subsec:AutomaticFeatures}, taking as a baseline the method proposed by Opara et al. \cite{Opara2020} in 2020.

\section*{Acknowledgment}
This work was supported by the framework agreement between the University of Le\'on and INCIBE (Spanish National Cybersecurity Institute) under Addendum 01.

\medskip

\bibliography{references.bib} 

\begin{thebibliography}{10}

\bibitem{Anti-PhishingWorkingGroup2020}
{Anti-Phishing Working Group}.
\newblock {Phishing Activity Trends Report 3 Quarter}.
\newblock {\em Most}, (March):1--12, 2020.
\newblock Available at
  \url{https://docs.apwg.org/reports/apwg_trends_report_q3_2020.pdf} Accessed:
  Jan.~14, 2021.

\bibitem{Jain2016}
Ankit~Kumar Jain and B.~B. Gupta.
\newblock {A novel approach to protect against phishing attacks at client side
  using auto-updated white-list}.
\newblock {\em Eurasip Journal on Information Security}, 2016(1):1--11, dec
  2016.

\bibitem{Prakash2010}
Pawan Prakash, Manish Kumar, Ramana {Rao Kompella}, and Minaxi Gupta.
\newblock {PhishNet: Predictive blacklisting to detect phishing attacks}.
\newblock In {\em Proceedings - IEEE INFOCOM}, 2010.

\bibitem{Sahingoz2019}
Ozgur~Koray Sahingoz, Ebubekir Buber, Onder Demir, and Banu Diri.
\newblock {Machine learning based phishing detection from URLs}.
\newblock {\em Expert Systems with Applications}, 117:345--357, mar 2019.

\bibitem{Moghimi2016}
Mahmood Moghimi and Ali~Yazdian Varjani.
\newblock {New rule-based phishing detection method}.
\newblock {\em Expert Systems with Applications}, 53:231--242, jul 2016.

\bibitem{Somesha2020}
M.~Somesha, Alwyn~Roshan Pais, Routhu~Srinivasa Rao, and Vikram~Singh Rathour.
\newblock {Efficient deep learning techniques for the detection of phishing
  websites}.
\newblock {\em Sadhana - Academy Proceedings in Engineering Sciences},
  45(1):1--18, dec 2020.

\bibitem{Zhu2019}
Erzhou Zhu, Yuyang Chen, Chengcheng Ye, Xuejun Li, and Feng Liu.
\newblock {OFS-NN: An Effective Phishing Websites Detection Model Based on
  Optimal Feature Selection and Neural Network}.
\newblock {\em IEEE Access}, 7:73271--73284, 2019.

\bibitem{Paniagua19}
M.~S{\'a}nchez-Paniagua, E.~Fidalgo, V.~Gonz{\'a}lez-Castro, and E.~Alegre.
\newblock Impact of current phishing strategies in machine learning models for
  phishing detection.
\newblock In {\'A}lvaro Herrero, Carlos Cambra, Daniel Urda, Javier Sedano,
  H{\'e}ctor Quinti{\'a}n, and Emilio Corchado, editors, {\em 13th
  International Conference on Computational Intelligence in Security for
  Information Systems (CISIS 2020)}, pages 87--96, Cham, 2021. Springer
  International Publishing.

\bibitem{Gangwar18}
Abhishek Gangwar, Eduardo Fidalgo, Enrique Alegre, and V{\'\i}ctor
  Gonz{\'a}lez-Castro.
\newblock Phishfingerprint: A practical approach for phishing web page identity
  retrieval based on visual cues.
\newblock In {\em International Conference of Applications of Intelligent
  Systems}, Las Palmas de Gran Canaria, 2018.

\bibitem{Opara2019}
Chidimma Opara, Bo~Wei, and Yingke Chen.
\newblock {HTMLPhish: Enabling Phishing Web Page Detection by Applying Deep
  Learning Techniques on HTML Analysis}.
\newblock {\em Proceedings of the International Joint Conference on Neural
  Networks}, aug 2019.

\bibitem{Opara2020}
Chidimma Opara, Yingke Chen, and Bo. Wei.
\newblock {Look Before You Leap: Detecting Phishing Web Pages by Exploiting Raw
  URL And HTML Characteristics}.
\newblock {\em arXiv preprint arXiv:2011.04412}, nov 2020.

\bibitem{Adebowale2019}
Moruf~A Adebowale, Khin~T Lwin, Erika Sanchez, and M~Alamgir Hossain.
\newblock Intelligent web-phishing detection and protection scheme using
  integrated features of images, frames and text.
\newblock {\em Expert Systems with Applications}, 115:300--313, 2019.

\bibitem{Mohammad2015}
Rami~M Mohammad, Fadi Thabtah, and Lee McCluskey.
\newblock Phishing websites features.
\newblock {\em School of Computing and Engineering, University of
  Huddersfield}, 2015.
\newblock Available at
  \url{http://archive.ics.uci.edu/ml/datasets/Phishing+Websites} Accessed:
  Jan.~14, 2021.

\bibitem{Yang2019}
Peng Yang, Guangzhen Zhao, and Peng Zeng.
\newblock {Phishing website detection based on multidimensional features driven
  by deep learning}.
\newblock {\em IEEE Access}, 7:15196--15209, 2019.

\bibitem{Rao2019}
Routhu~Srinivasa Rao and Alwyn~Roshan Pais.
\newblock {Detection of phishing websites using an efficient feature-based
  machine learning framework}.
\newblock {\em Neural Computing and Applications}, 31(8):3851--3873, aug 2019.

\bibitem{Ozker2020}
U{\u{g}}ur Ozker and Ozgur~Koray Sahingoz.
\newblock Content based phishing detection with machine learning.
\newblock In {\em 2020 International Conference on Electrical Engineering
  (ICEE)}, pages 1--6. IEEE, 2020.

\bibitem{Li2019}
Yukun Li, Zhenguo Yang, Xu~Chen, Huaping Yuan, and Wenyin Liu.
\newblock {A stacking model using URL and HTML features for phishing webpage
  detection}.
\newblock {\em Future Generation Computer Systems}, 94:27--39, may 2019.

\bibitem{Alotaibi2020}
Bandar Alotaibi and Munif Alotaibi.
\newblock {Consensus and majority vote feature selection methods and a
  detection technique for web phishing}.
\newblock {\em Journal of Ambient Intelligence and Humanized Computing}, 1:3,
  may 2020.

\bibitem{Tan2018}
Choon~Lin Tan.
\newblock Phishing dataset for machine learning: Feature evaluation.
\newblock {\em Mendeley Data}, 1:2018, 2018.

\end{thebibliography}

\newpage




\end{document}